# Phase Matched Plasmonic Transmission Lines for Cascaded Second-Harmonic Generation as a Pathway to Nonlinear Logic Circuits

Komal Gupta,[1] Anand Hegde,[1] Xiaofei Wu,[2] Jer-Shing Huang,[2,3,4,5] and Chen-Bin Huang[1,4,*]

## Affiliations

1 Institute of Photonics Technologies, National Tsing Hua University, Hsinchu, Taiwan.

2 Leibniz Institute of Photonic Technology, Albert-Einstein Strasse 9, Jena, Germany

3 Institute of Physical Chemistry and Abbe Center of Photonics, Friedrich-Schiller-Universität Jena, Helmholtzweg 4, Jena, Germany

4 Research Center for Applied Sciences, Academia Sinica, 128 Sec. 2, Academia Road, Taipei, Nangang, Taiwan

5 Department of Electrophysics, National Yang Ming Chiao Tung University, Hsinchu, Taiwan.

*Corresponding author: robin@ee.nthu.edu.tw

## Abstract

In nonlinear nanophotonics, cascaded second-harmonic generation (SHG) in pure plasmonic waveguides for sequential signal transformation and complex on-chip functionality remains a long-standing challenge. Precise phase matching becomes instrumental to achieve efficient SHG and enable true cascading of nonlinear processes. We experimentally demonstrate phase matching in SHG is achievable in a plasmonic system between two orthogonal modes. Accurate tuning of plasmonic two-wire transmission-line (TWTL) design parameters result in SHG from the antisymmetric excitation mode being approximately 15 times stronger than that from the symmetric excitation mode, greatly raising the conversion efficiency to 0.021%. Simultaneously, phase matching extends our TWTL operational length up to 18 μm. Based on the improved efficiency and operational length, we demonstrate the feasibility of cascading multiple plasmonic waveguides. We further realize a nonlinear AND logic operation using a single 18 μm long waveguide. These results underscore the potential of phase matched plasmonic TWTLs for compact, efficient, and scalable nonlinear optical circuitry.

## Introduction

Programmable nonlinear photonic processors that perform on-chip frequency conversion and sequential signal processing are important foundations toward adaptable optical computing (*1, 2*), yet most recent demonstrations use multi-material platforms that add fabrication complexity and limit integration density (*3, 4*). These functions are often implemented via second-harmonic generation (SHG), which doubles the fundamental frequency and enables wavelength translation in compact devices. In various circuit functions, cascading different stages are crucial. Such nonlinear cascading requires high conversion efficiency and a long effective interaction length. Therefore, phase matching becomes crucial which enables robust cascading and more elaborate circuit layouts (*5, 6*). Past works have demonstrated modal and quasi-phase matching in thin-film ZnS (*7*), LiNbO$_3$ (*8*), III–V waveguides (*9, 10*), and other semiconductor and metasurface platforms (*11, 12*), enhancing SHG through dispersion engineering. These diverse approaches illustrate the versatility of phase matching across material systems and device architectures, underscoring its central role in advancing integrated nonlinear photonics.

In plasmonic platforms, most demonstrations of SHG employed localized surface plasmon resonances (LSPRs). Nanoscale confinement relaxes the phase matching requirement but directs the harmonic generations into free space with wide angular spread (*13-19*). This limits their use for on-chip routing. In contrast, propagating surface plasmon polaritons support guided emission with polarization control and directionality. They are attractive for on-chip routing but phase matching is difficult because of large effective index and strong dispersion which shorten the coherence and operational length (*20, 21*). Hybrid plasmonic–photonic waveguides match plasmonic and dielectric modes to recover momentum. These designs improve coupling into guided channels but require multi-material processing involving complicated fabrication procedures and precise alignment (*22-27*). All metal plasmonic waveguides offer single material fabrication and deep subwavelength confinement. The only missing piece is phase matching with guided SHG output in pure metal geometries (*28*). Without phase matching, the coherence length and conversion efficiency both remain low, hindering the development of advanced photonic circuitry.

Firstly, we demonstrate cross-modal phase matching between the fundamental frequency mode (FFM) at ω and a guided second harmonic (SH) mode at 2ω in a plasmonic two-wire transmission line (TWTL). This configuration enables efficient SHG in a compact all-metal geometry and avoids hybrid materials or periodic modulation. Further, we show nonlinear cascadability and logic operation. These capabilities were previously experimentally demonstrated only in metal–insulator–metal (MIM) waveguides (*29*) or single silver nanowires (*30*) and even then limited to linear logic operations. Our approach leverages the surface-mediated second-order response of gold and geometric dispersion control. It delivers internal SHG efficiencies that approach values reported for hybrid plasmonic platforms in the range $\sim 10^{-2}$ to $10^{-4}$ (*23*). It exceeds typical all-metal values near $10^{-5}$ (*31*). This all-metal platform establishes a route to phase matched plasmonic frequency conversion for on-chip logic and wavelength translation. It also opens opportunities for quantum-light generation via spontaneous parametric down-conversion (SPDC) and for scalable integrated photonics with further gains possible using two-dimensional materials or gain media (*32-34*).

## Results

### Phase matching reverses SHG mode contrast

The TWTL device consists of two identical Au nanowires connected by an incoupling link antenna and a mode detector (single-stub termination) fabricated on a $SiO_2$ substrate. The structure is designed to support two distinct propagation modes: symmetric and antisymmetric, depending on the polarization state of the incident light (*21*). When the electric field vector of the incident light is parallel to the nanowires ($\mathbf{E}_\parallel$), the symmetric mode is selectively excited, whereas a perpendicular field orientation ($\mathbf{E}_\perp$) excites the antisymmetric mode. In either excitation condition, a local nonlinear polarization vector $\mathbf{P}^{(2)}(2\omega)$, is generated, resulting in SH emission into the symmetric mode according to $\mathbf{P}^{(2)} \propto \mathbf{E}_{in}^2$.

Fig. 1 highlights the contrast in SHG between a previously reported non-phase matched TWTL and the current phase matched plasmonic design. Fig. 1A schematically illustrates the operating principle of the TWTL when symmetric and antisymmetric FFM(s) are excited, showing the propagation of SPPs at both the FM and SH frequencies. The cross-sectional field distribution obtained through numerical modeling reveals the spatial confinement of the FFM and SH modes in a non-phase matched configuration. In the absence of phase matching, previous results demonstrate that the symmetric FFM produces a symmetric SH mode with an intensity 5.5 times greater than that generated by the antisymmetric FFM (*21*). The inset of Fig. 1A presents the device parameters, including nanowire width (*w*), wire thickness (*t*), and gap between the two wires (*g*). Phase matching is achieved by tailoring these parameters to modify the effective refractive indices of the interacting modes, ensuring matched phase velocities and thereby enhancing nonlinear conversion efficiency.

In the optimized phase matched configuration, the SH intensity distribution is inverted: the antisymmetric FFM now produces a stronger SH signal than the symmetric FFM. Fig. 1B displays the cross-sectional modal profiles of the excited symmetric and antisymmetric FFMs and the corresponding generated symmetric SH mode in the phase matched device. For the antisymmetric mode, a phase difference $\Delta\phi = \pi$ between the FM and SH waves is observed at a given cross-section, suppressing modal phase overlap. Efficient SH generation is achieved by controlling the nanowire gap, which directly influences the effective refractive index $n_{eff} = n' + in''$ where $n'$ and $n''$ represent the real and imaginary components, respectively. The phase mismatch $\Delta k$ is determined using Equation (1) where, $n_\omega$ and $n_{2\omega}$ correspond to the real parts of the refractive indices at the FM and SH frequencies, respectively. Precise geometric optimization minimizes $\Delta k$ to near zero, resulting in dominance of the antisymmetric mode in SHG emission and evidencing the governing influence of phase matching on nonlinear mode interactions in plasmonic waveguides.

$$\Delta k = \frac{4\pi}{\lambda_{FM}} (n_{2\omega} - n_\omega) \tag{1}$$

**Geometric control of phase matching**

The simulations reveal that precise geometric control of the TWTL offers a direct and powerful route to phase matching between the antisymmetric FM and the symmetric SH mode. The inter-wire gap (*g*) emerges as the dominant parameter governing the effective refractive index of the antisymmetric FM, while leaving the symmetric SH mode largely unaffected. In the symmetric FM, a uniform charge distribution produces in-phase electric fields across the gap, whereas in the antisymmetric FM, uneven charge distribution induces a fixed phase shift of $\Delta\phi = \pi$, yielding out-of-phase fields. Although the phase relationships remain invariant along the nanowire length, variations in *g* produce substantial shifts in the modal indices, directly modifying propagation velocities and nonlinear interaction strength.

The influence of nanowire width (*w*) and thickness (*t*) is secondary but non-negligible. Width variation has minimal impact on field confinement, yet excessive *w* excites higher-order modes that increase loss. Increasing *t* in the SH mode induces leaky-mode behavior due to activation of higher-order modes (Supplementary Information S1).

In the non-phase matched configuration ($w$= 140 nm, $g$= 100nm, and $t$= 60nm), dispersion curves (Fig. 2A) exhibit a pronounced mismatch between the antisymmetric FM and symmetric SH mode, effectively suppressing nonlinear conversion. Systematic variation of $g$ (Fig. 2B) shows that the antisymmetric FM index decreases sharply with increasing gap, intersecting with the nearly constant index of the symmetric SH mode at $g$=70nm, thereby fulfilling the phase matching condition.

Gap tuning delivers strong modal index control but is highly sensitive, making it less suited for precise adjustment. Width variation (Fig. 2C) offers a finer, more controllable tuning mechanism, enabling targeted index alignment without compromising field enhancement. The optimized dispersion relation (Fig. 2D) confirms that combined tuning of $g$ and $w$ achieves exact intersection of the FM and SH mode curves, validating full phase matching. This engineered dispersion control provides a deterministic strategy for inverting SHG mode contrast and maximizing nonlinear efficiency in plasmonic waveguides.

**Experimental validation**

Previous studies show stronger SHG from symmetric mode excitation than from antisymmetric excitation(*21, 24*). Here, phase matching is engineered to reverse this contrast by enhancing nonlinear coupling in the antisymmetric mode. Numerical optimization of TWTL length under phase matched and non-phase matched conditions (Supplementary Information S2) guides the fabrication of devices with varying gap sizes.

Fig. 3A shows the experimental setup: the antenna is excited from the substrate side with a 1550 nm femtosecond laser, and SH emission is collected through an analyzer parallel ($\theta_A = 0°$) to the waveguide axis and imaged with an EMCCD. Fig. 3B presents the measured SHG outputs. Under phase matched conditions, antisymmetric excitation yields a significantly higher SH signal than symmetric excitation, confirming the predicted reversal of SHG mode contrast.

Quantitative analysis is performed by comparing simulated SH energies with experimentally measured SH photon counts ($n_{SH}$) across a range of gap sizes under antisymmetric excitation. As shown in Fig. 3C, both simulation and experiment exhibit a pronounced SHG maximum at the phase matched gap, confirming the sensitivity of SH generation to effective index matching, consistent with the trends in Fig. 2B. The nonlinear nature of the process is verified in Fig. 3D, where $n_{SH}$ scales quadratically with input power. At optimal phase matching, antisymmetric excitation produces a 15.5-fold enhancement in SHG compared to symmetric excitation.

To assess SHG conversion efficiency ($\eta$), the incoupling efficiency ($\eta_{in}$) and the outcoupling efficiency ($\eta_{out}$) of the optimized device are numerically calculated. Using the calculated power ($P_0$) at a distance $l_d$ from the excitation port and the input power ($P_{in}$), ($\eta_{in}$) and ($\eta_{out}$) efficiencies are determined from Equation (2). The overall SHG conversion efficiency is then evaluated using Equation (3), where, $N_{SH}$ and $N_{FM}$ are the number of photons counts per second for SH and FM, respectively.

$$\eta_{\frac{in}{out}} = \frac{P_0}{P_{in}} \left( e^{4\pi n'' l_d / \lambda_{FM}} \right) \qquad (2)$$

$$\eta = \frac{N_{SH}\, \eta_{out}}{N_{FM}\, \eta_{in}} \qquad (3)$$

The phase matched TWTL achieves an incoupling efficiency of 6.024% and an outcoupling efficiency of 51.6%. Based on these values, the experimental SHG conversion efficiency is calculated as 0.021%. For an input FM power of 6.73×10⁻¹⁰ W/cm², the estimated SH power reached 1.41×10⁻¹³ W/cm².

**SHG propagation and operational length**

The influence of phase matching on the propagation length of the TWTL is evaluated by tracking the evolution of SH intensity with waveguide length. In an ideal, lossless system, the coherence length is given by $l_c = \pi/\Delta k$, defining the distance over which the FM and SH waves remain in phase and interact constructively. In plasmonic waveguides, however, intrinsic ohmic losses attenuate the propagating surface plasmons, reducing the effective nonlinear interaction length well below this ideal value.

To experimentally probe this effect, a series of TWTL devices is fabricated with lengths ranging from 8 μm to 18 μm. Fig. 4A presents representative scanning electron microscopy (SEM) images of the fabricated devices. The structures are excited with antisymmetric mode, and the emitted SH signals are collected with an EMCCD through an analyzer ($\theta_A = 0°$). In the experimental images, faint scattering is observed near the center of some waveguides, arising from localized surface plasmons excited by residual nanoparticles left from fabrication. Numerical simulations confirm that these scattering centers contribute negligibly to the measured SHG (Supplementary Information S3).

The dependence of SHG on propagation length is shown in Fig. 4B, which compares simulated SH energies and experimentally measured normalized $n_{SH}$ for both symmetric and antisymmetric excitation. Under phase matched conditions, the antisymmetric mode exhibits a steady increase in SH intensity, reflecting continuous and efficient energy transfer from the FM to the SH mode along the propagation path. By contrast, the symmetric mode, which is phase-mismatched, displays a periodic modulation in SH output. This oscillatory behavior arises from alternating constructive and destructive interference between the FM and SH waves, limiting the net conversion efficiency over extended distances.

These measurements demonstrate that phase matched antisymmetric modes can sustain efficient SHG over lengths up to 18 μm without noticeable saturation. From the measured dispersion relation, the group-velocity dispersion parameters are extracted using Equation (4) and (5), yielding $D_\lambda = -3744\, ps/km/nm$ and $\beta_2 = 1.2 \times 10^{-9} ps^2/nm$. Although group velocity dispersion may affect longer SH pulses, plasmonic losses dominate, leading to intensity decay rather than temporal broadening (Supplementary Information S4).

$$D_\lambda = \frac{-\lambda}{c} \frac{d^2 n_{eff}}{d\lambda^2} \qquad (4)$$

$$\beta_2 = \frac{-\lambda^2}{2\pi c} D_\lambda \qquad (5)$$

**Cascaded SHG and polarization-controlled logic operations**

Cascading in nonlinear optics allows the output of one nonlinear process to serve as the input to another, enabling sequential signal transformation and complex functionality. In the phase-matched TWTL, the experimentally achieved enhancement in operational length provides the coherent interaction needed for such cascading. This potential is examined by testing whether the generated SH signal can act as a coherent source for a subsequent nonlinear stage. Fig. 5A shows two phase matched TWTL devices fabricated in a parallel configuration for sequential operation. Because SHG occurs in the symmetric mode, the devices are aligned to maintain mode continuity between stages. In the first stage, antisymmetric FM excitation generates a symmetric SH output of intensity $I_1$. This SH signal is then directly coupled into the second TWTL, where it propagates and undergoes an additional SHG process, yielding output intensity $I_2$ as shown in Fig. 5B. The comparison in Fig. 5C between normalized $I_1$ and $I_2$ confirms that the SH generated in the first stage is sufficiently strong and coherent to drive a second nonlinear interaction. These results establish the feasibility of cascading plasmonic nonlinear stages, a capability relevant to advanced functionalities such as SPDC and nonlinear optical logic.

Nonlinear plasmonic logic gates can map optical input states to distinct nonlinear outputs. A lone unit of TWTL from previously demonstrated cascaded device (Fig. 5A) is capable of independently performing nonlinear logic. Leveraging the phase matched TWTL, a polarization-controlled nonlinear logic gate is demonstrated using an 18 µm-long device. Fig. 6A shows the SEM image of the fabricated structure with the excitation port indicated. Two logical inputs, A and B, are defined by the FM intensities coupled into each nanowire, while SH outputs $I_0$ and $I$ are measured through analyzers oriented perpendicular or parallel ($\theta_A = 90° \text{ or } 0°$) to the waveguide axis. Under ±45° excitation (Fig. 6B), the FM field predominantly couples into a single nanowire, corresponding to input states (1,0) or (0,1), and yields negligible SH at output $I$, defined as Boolean 0 in the SH domain $E(2\omega)$. In contrast, 90° excitation (Fig. 6C) launches equal FM intensities into both nanowires, producing SH signal at $I$ defined as Boolean 1.

The logic mapping is summarized in Fig. 6D as an AND operation, with SH generation occurring only when both inputs are active. To demonstrate tunability, the analyzer angle ($\theta_A$) is varied while maintaining antisymmetric FM excitation. The measured modulation in SH output (Fig. 6E) confirms continuous control over the nonlinear response through polarization-defined inputs and detection geometry. This polarization-encoded operation illustrates how phase matched plasmonic waveguides can integrate reconfigurable nonlinear logic functions into on-chip photonic architectures.

**Discussion**

This work reveals that cross-modal phase matching in pure intrinsic plasmonic TWTL waveguides unlocks nonlinear cascadability and logic. By achieving previously elusive modal phase matching through precise geometric control, we demonstrate a substantial enhancement in SHG efficiency. Tuning the nanowire gap and width, the effective indices of the antisymmetric FM and symmetric SH mode are matched, facilitating efficient nonlinear coupling. This design not only reverses the conventional trend of symmetric-mode dominance in SHG by 15-fold but also delivers a measurable efficiency of 0.021% and extends the effective operational length to 18µm, significantly beyond that of previous non-phase matched plasmonic implementations.

The results demonstrate that the generated SHG signal in a phase matched TWTL is sufficiently strong and coherent to drive a second nonlinear stage, confirming the feasibility of cascaded nonlinear plasmonic processing. This multi-stage configuration highlights the potential for implementing frequency-conversion chains and even SPDC, a key process for generating entangled photons in quantum optics. In parallel, polarization control is harnessed to implement a basic nonlinear logic operation. By encoding inputs in polarization states and detecting SH outputs through a polarization-sensitive analyzer, an optical AND gate is realized within a single 18 μm long TWTL. Such polarization-encoded functionality establishes a framework for scalable, logic-enabled nonlinear photonic circuits.

While these demonstrations mark a substantial advance, the present devices face limitations in conversion efficiency due to intrinsic plasmonic losses and fabrication-induced imperfections such as surface roughness and residual nanoparticles. These effects modestly reduce SH signal strength and coherence, particularly over longer propagation lengths. Furthermore, the logic implementation relies on external polarization control and detection, which may present integration challenges. Both limitations represent well-defined engineering problems that can be addressed through improved nanofabrication and the development of on-chip polarization and phase control.

Looking forward, integrating two-dimensional materials or optical gain media into the plasmonic architecture not only offers a promising route to further enhance nonlinear efficiency but also unlocks greater degree of freedom for light–matter interactions. Such control would open new avenues for actively routing nonlinear signals and implementing more sophisticated logic operations directly on-chip. These advances will accelerate the development of practical, phase matched, and cascaded nonlinear elements for integrated photonics. Overall, this work lays a robust foundation for compact, reconfigurable nonlinear photonic systems with broad applicability in signal processing, on-chip logic, and quantum information technologies.

## Materials and Methods

### Numerical Simulations

We performed all numerical simulations using Ansys Lumerical FDTD and FDE solvers. We designed the device using the optical constants of gold from Johnson and Christy's material model (*35*). We excited the TWTL using a 1550 nm laser source with a 56fs pulse duration.

### Device Fabrication

We deposited a 60 nm-thick Au film onto the substrate using radio frequency (RF) sputtering and fabricated the TWTL devices using a Focused Ion Beam-Scanning Electron Microscope (FIB-SEM) system.

**Acknowledgments**

**Funding:** This work was supported by the National Science and Technology Council, Taiwan under Grant 113-2221-E-007-066-MY2 (C.-B. H.)

J.-S. H. acknowledges the support from DFG via projects 398816777 (SFB 1375) and 437527638 (IRTG 2675).

**Author contributions:**

Conceptualization and Methodology: KG, CBH

Fabrication: KG

Experimentation: KG, AH

Supervision: CBH, JSH, XW

Writing—original draft: KG, AH

Writing—review & editing: KG, AH, CBH, JSH

**Competing interests:** Authors declare that they have no competing interests.

**Data and materials availability:** All data supporting the findings of this study are available within the article and its supplementary materials. Additional raw data and materials used in this research are available from the corresponding author upon reasonable request.

**Figures**

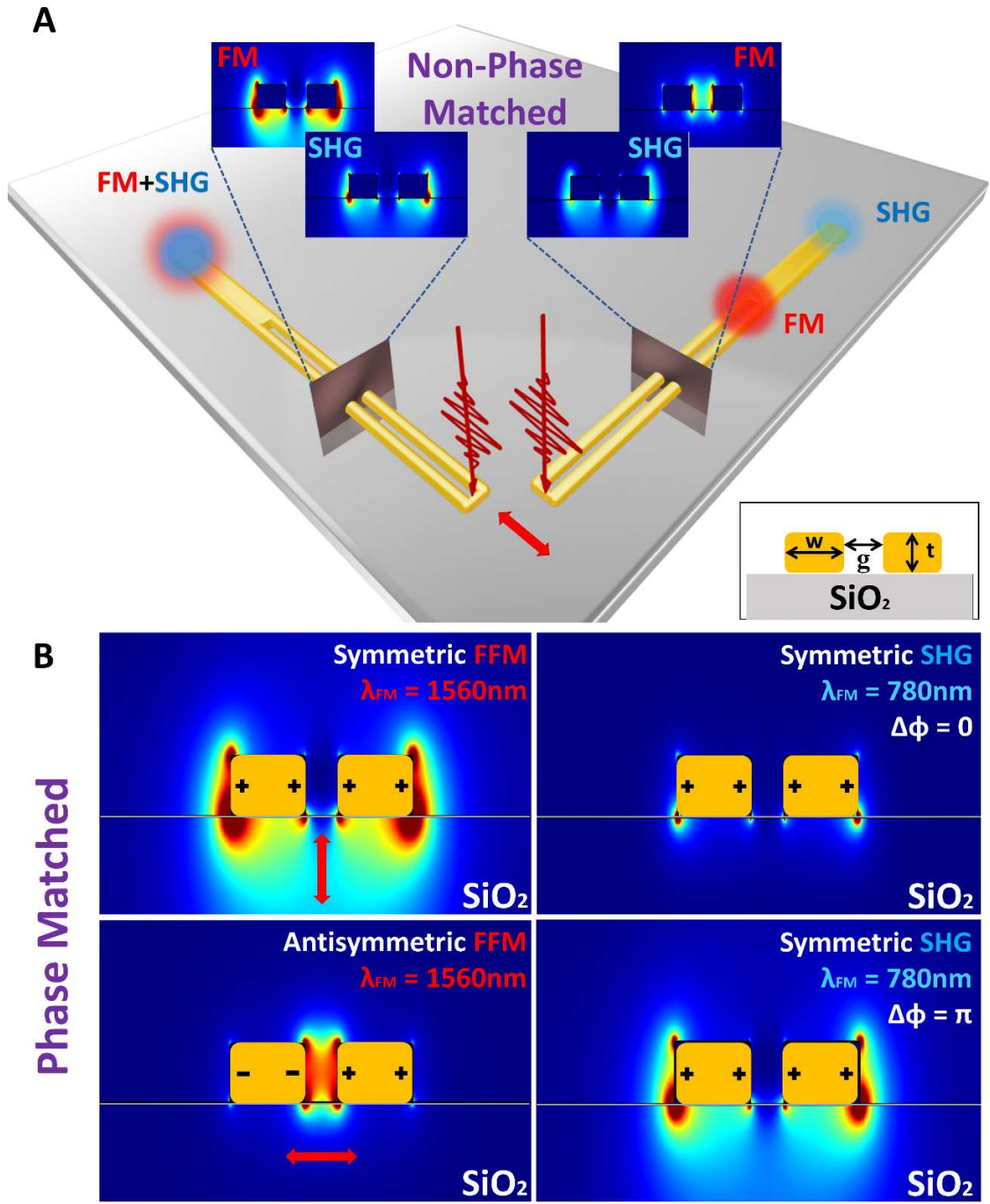

**Fig. 1. Working principle and modal analysis. (A)** Illustrates the excitation of the Au-TWTL with symmetric and antisymmetric FFM and propagating SH symmetric mode, respectively. The crossectional view presents the electric field confinement of the propagating FFM and SH modes for non-phase matched device, where the SH intensity generated from the symmetric FFM is stronger than that from the antisymmetric FFM. The inset provides the notation of designing parameters. **(B)** Simulated crossectional view of electric field confinement and charge distribution for excited FFM and generated SH modes for a phase matched device, showing that the SH intensity resulting from the antisymmetric FFM surpasses that of the symmetric FFM.

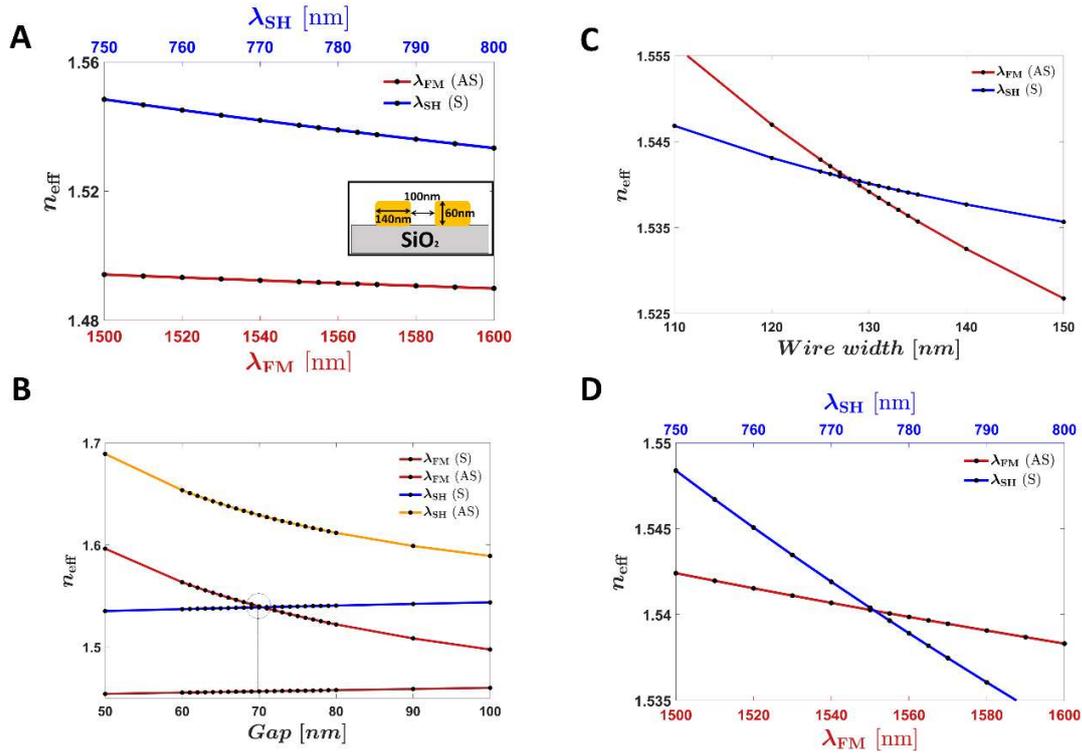

**Fig. 2. Dispersion relation and geometric tuning.** Numerically calculated (**A**) dispersion curve for antisymmetric FM and symmetric SH for non-phase matched device. The variation of effective refractive index with respect to the (**B**) gap distance, (**C**) wire width after fixing the thickness to 60nm. (**D**) Dispersion curve for antisymmetric FM and symmetric SH for phase matched device after tuning the designing parameters.

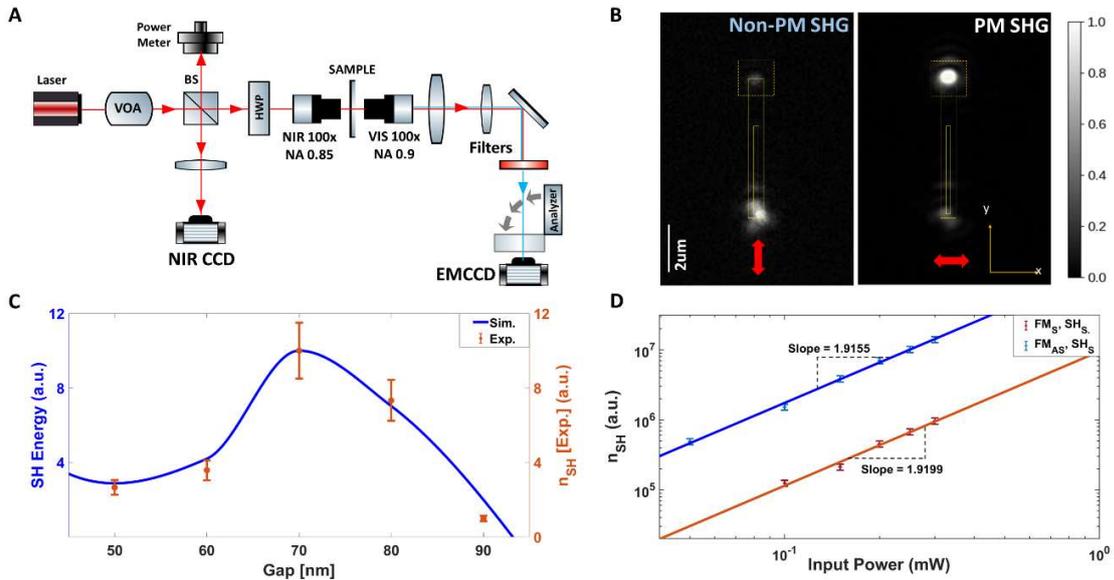

**Fig. 3. Experimental SHG contrast inversion by phase matching.** (**A**) Schematic of the experiment setup. (**B**) Experimental demonstration of SH propagating modes in the phase matched device for symmetric and antisymmetric excitations, respectively. (**C**) Simulated SH energy and experimental $n_{SH}$ for different gap sizes for antisymmetric excitation. (**D**)

$n_{SH}$ from the phase matched antisymmetric excitation vs non-phase matched symmetric excitation with varying input power in the same device.

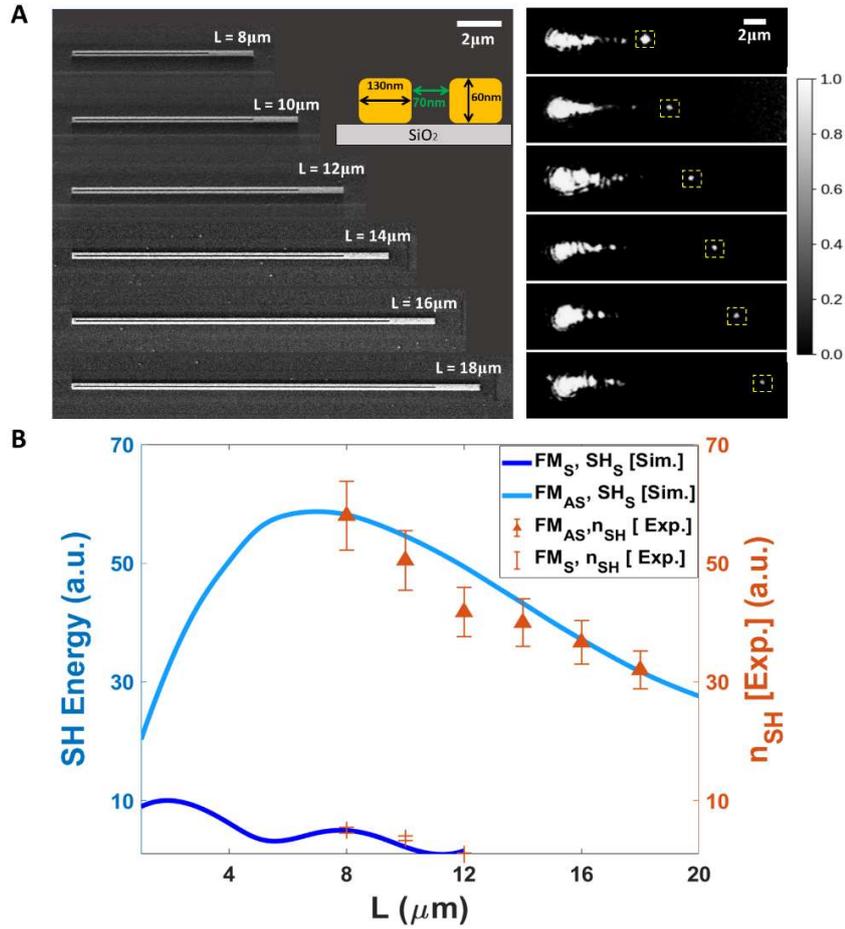

**Fig. 4. SHG propagation length in TWTL.** (**A**) Shows the SEM images of the TWTL devices of length 8um to 18um, and the SH signal obtained from the EMCCD for antisymmetric mode of excitation ($\theta_A$= parallel to the device orientation). (**B**) Numerically and Experimentally calculated SH energy and $n_{SH}$ variation with respect to the length of the TWTL comparing the excitation from antisymmetric and symmetric modes.

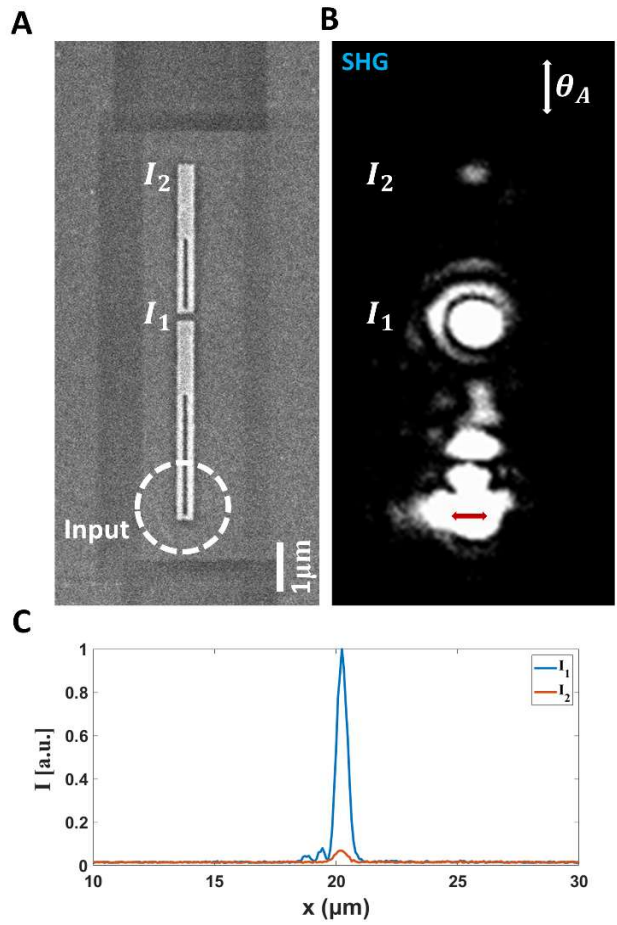

**Fig. 5. Cascaded SHG in phase matched TWTL.** (**A**) SEM of two phase matched TWTLs in a parallel cascaded layout. Antisymmetric FM excitation in the first stage generates a symmetric SH signal ($I_1$) that is directly coupled into the second stage. (**B**) EMCCD image of SH emission from both stages ($\theta_A = 0°$). (**C**) Normalized SH intensities from the first $I_1$ and second $I_2$ stages, confirming that the generated SH is strong and coherent enough to drive a subsequent nonlinear process.

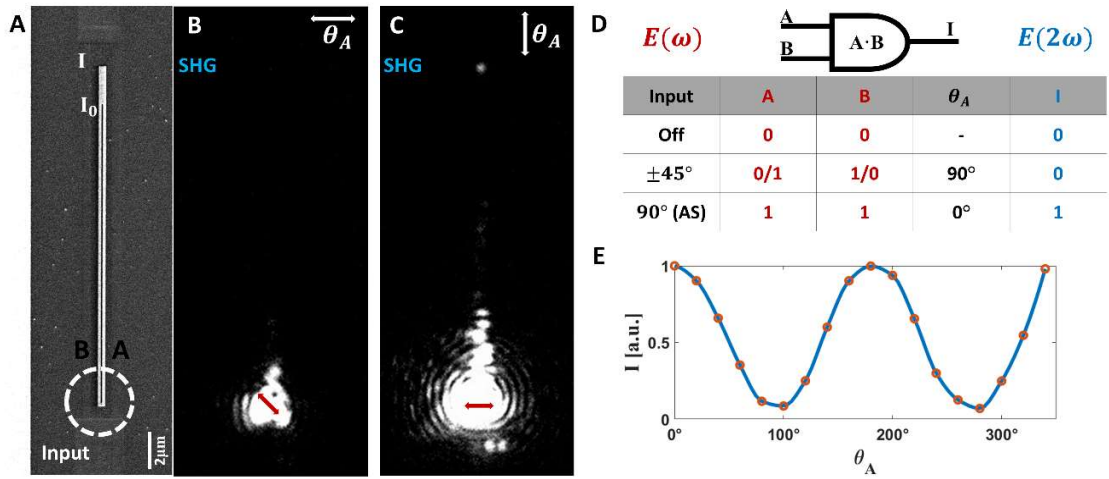

**Fig. 6. Polarization controlled nonlinear logic operation.** (**A**) SEM of an 18 μm-long phase matched TWTL showing the excitation port, linear input A and B, and SH output $I_0$ and $I$. (**B**) EMCCD SHG image for ±45° polarization excitations, where the FM couples predominantly into a single nanowire (logic inputs (0,1) or (1,0)), producing negligible SH output $I$ (Boolean 0) when $\theta_A = 90°$. (**C**) EMCCD SHG image for 90° polarization excitations, where equal FM intensities are launched into both nanowires (logic input (1,1)), producing a strong SH output $I$ (Boolean 1) when $\theta_A = 0°$. (**D**) AND logic truth table summarizing the SH output states, with Boolean 1 occurring only for antisymmetric FM excitation. (**E**) SHG output intensity ($I$) as a function of analyzer angle ($\theta_A$) under antisymmetric excitation, showing continuous polarization-based control of nonlinear logic states.

**Supplementary Materials**

S1. Impact of varying device thickness
S2. Device length optimization
S3. SH spectrum for antisymmetric and symmetric emission
S4. Dispersion Analysis

# Supplementary Materials for

# Phase Matched Plasmonic Transmission Lines for Cascaded Second-Harmonic Generation as a Pathway to Nonlinear Logic Circuits

Komal Gupta *et al.*

*Corresponding author: Chen-Bin Huang, robin@ee.nthu.edu.tw

**This PDF file includes:**

Supplementary Text
Figs. S1 to S4
Table S1
References (21, 35 to 36)

## S1. Impact of varying device thickness

To investigate the effect of device thickness on fundamental (FM) and second harmonic (SH) scattering intensities, we use FDTD and FDE simulations in Lumerical. The optical properties of the waveguide consisting of gold are taken as interpolated data from Johnson and Christy (35). For each thickness ($t$), we first optimize the wire width ($w$) and gap ($g$) to achieve phase matching. Table 1 shows the optimized parameters and the effective index mismatch ($\Delta n$) between the FM and SH modes.

After obtaining the optimized parameters, we study the variation of SH intensity with respect to the wire length to determine the optimal length. We then analyze the far-field scattering of the FM and SH signals for the optimized structure length. Fig. S1 shows the far-field scattering patterns of the fundamental (FM) and second harmonic (SH) signals for structures with optimized length and phase-matching parameters. The images demonstrate how scattering intensity and mode contributions change with increasing device thickness. For the optimized devices, we observe that the FM scattering spot becomes more intense as the thickness increases. For the SH signal, the TM scattering spot intensifies as the thickness increases from 40 nm to 60 nm. At 70 nm thickness, the TM scattering is comparable to that at 60 nm, but the contributions from localized surface plasmon (LSP) modes and higher-order modes are stronger. At 80 nm thickness, the SH TM scattering spot diminishes. Accordingly, a thickness of 60 nm is selected.

## S2. Device length optimization

We numerically excite the TWTL device using a 1550 nm, 56 fs laser source and monitor the emission of the SH signal along the length of an extended TWTL structure, recording the SH output at intervals of 1 μm. This allows us to analyze how the SH generation evolves as the interaction length increases. Fig. S2 shows the SH energy as a function of device length for different gap sizes. From these trends, we identify the optimal device length for each gap size as the point where the SH energy reaches its maximum, indicating the most efficient nonlinear interaction for that configuration.

## S3. SH spectrum for antisymmetric and symmetric emission

Numerical simulations and previous studies (21) show that the contribution to SHG from the gap region is minimal, as illustrated in Fig. S3. To confirm this, we analyze the SH spectra at the inner and outer ends of the mode detector. The results indicate that the SH signal primarily originates from the symmetric guided modes detected at the outer end, rather than from localized interactions in the gap region. Fig. S3 shows distinct spectral signatures for (A) antisymmetric emission at the inner end, which contains only the fundamental mode (FM), and (B) symmetric emission at the outer end, which includes the SH signal along with a weaker FM component. These results further support the conclusion of negligible SHG contribution from the gap region.

## S4. Dispersion Analysis

Dispersion in the TWTL is characterized by the effective refractive index ($n_{eff}$) of the plasmonic mode as a function of wavelength ($\lambda$), described by the derivative $dn_{eff}/d\lambda$ (36). Fig. S4(A) shows the $dn/d\lambda$ plot for both the FM and the SH, derived from the data presented in Fig. 2D.

The blue curve in Fig. S4(A) represents the variation of $dn/d\lambda$ with respect to the FM wavelength. This curve is nearly a straight line with an almost constant value, indicating that the group index remains nearly constant. As a result, the group velocity is effectively constant, and the group velocity dispersion (GVD) is minimal. Although higher-order dispersion terms are not entirely absent, they are sufficiently small that their impact on the pulse shape and spectral components is negligible over practical propagation distances. The red curve in Fig. S4(A) shows the variation of $dn/d\lambda$ with respect to the SH wavelength. This curve has a slightly positive slope, indicating the presence of weak GVD. This weak dispersion can introduce gradual pulse broadening over longer propagation distances, but the effect remains minimal in short devices. Fig. S4(B) presents the numerically calculated SH spectra for different device lengths, confirming that the impact of dispersion on the SH signal is negligible. While the weak dispersion may cause minor pulse shape distortion and slight signal degradation at the SH wavelength, the dominant effect remains amplitude decay due to intrinsic plasmonic losses.

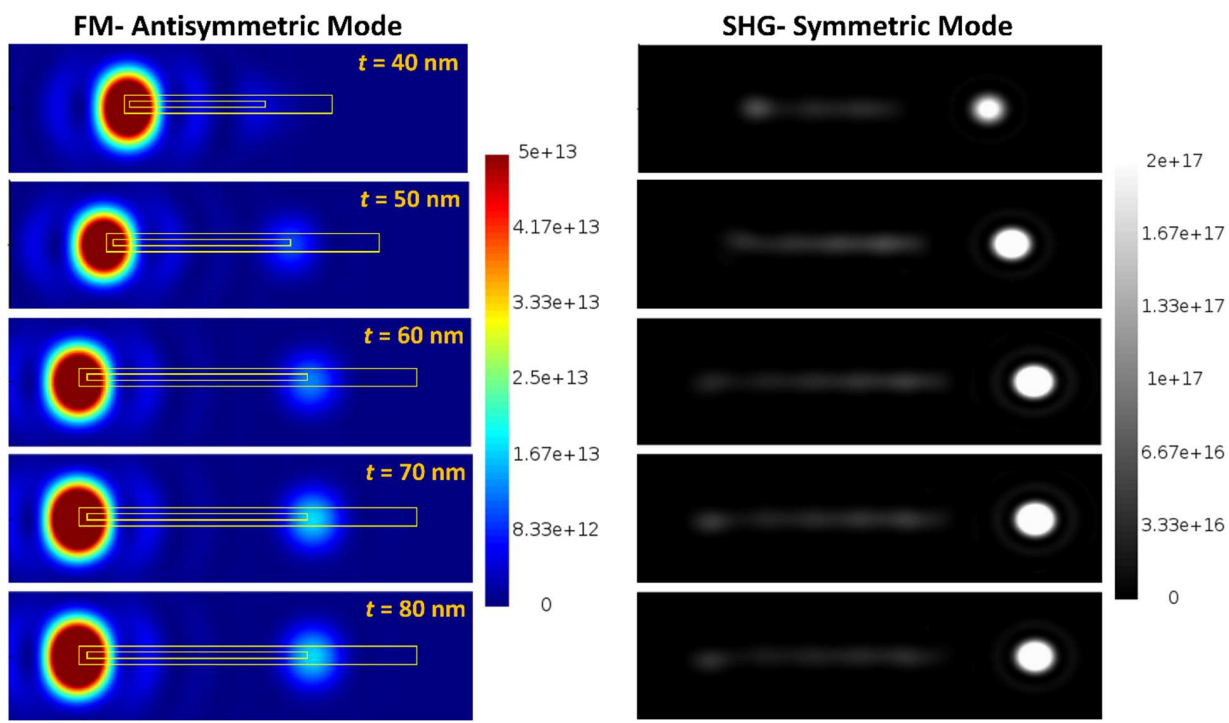

**Fig. S1. Impact of varying device thickness.** Far-field scattering of FM and SH signals for varying thicknesses, illustrating changes in scattering intensity and mode contributions

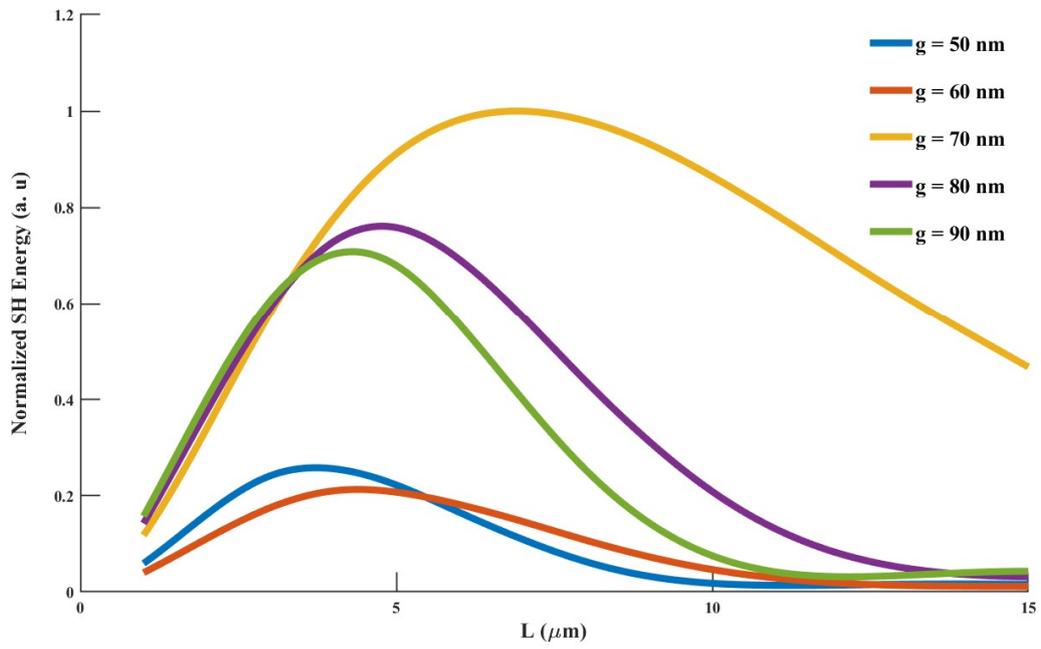

**Fig. S2**. **Device length optimization.** Second harmonic energy as a function of device length for various gap sizes

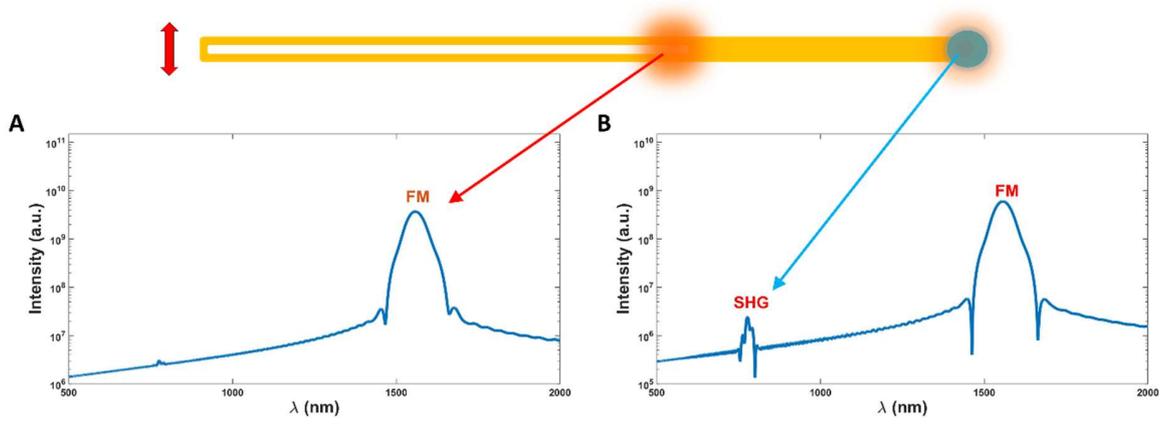

**Fig. S3. SH spectrum analysis.** (A) Antisymmetric spectrum at the inner end showing only FM. (B) Symmetric spectrum at the outer end showing SH and weaker FM.

21 | P a g e

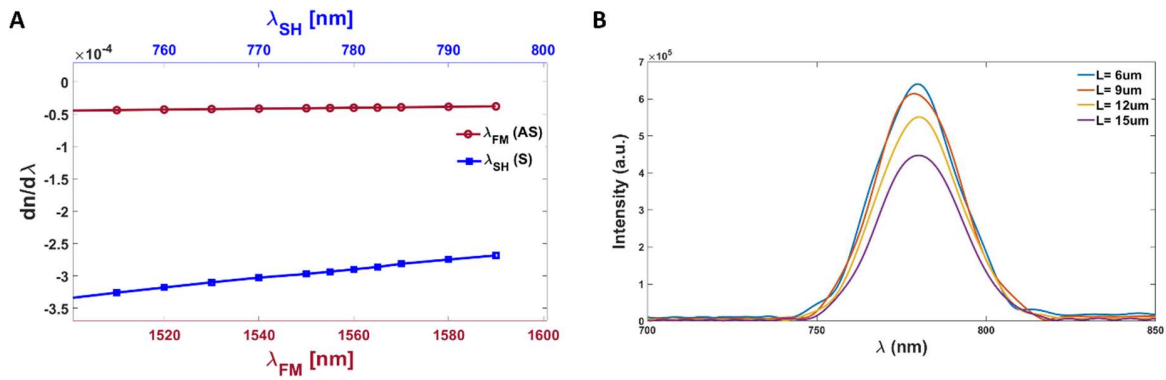

**Fig. S4. Dispersion analysis.** (A) dn/dλ for the fundamental mode (blue) and second harmonic (red), showing negligible dispersion for FM and weak dispersion for SH. (B) Simulated SH spectra for different device lengths, indicating negligible dispersion effects and dominant amplitude decay.

**Table S1.** Optimized wire width and gap values for each thickness, with corresponding effective index mismatch ($\Delta n$) between the fundamental and second harmonic modes

| $t$ [nm] | $g$ [nm] | $w$ [nm] | $\Delta n = |n_{2\omega} - n_{\omega}|$ |
|---|---|---|---|
| 40 | 80 | 120 | 0.001261 |
| 50 | 75 | 130 | 0.000245 |
| 60 | 70 | 130 | 0.000952 |
| 70 | 65 | 130 | 0.002079 |
| 80 | 60 | 130 | 0.003535 |
| 90 | 57 | 140 | 0.001473 |